# Automatic Segmentation of Choroid Layer in EDI OCT Images Using Graph Theory in Neutrosophic Space

**Bahareh Salafian[a,*], Rahele Kafieh[b], Abdolreza Rashno[a], Mohsen Pourazizi[c,**], Saeid Sadri[a]**

[a] Department of Electrical and Computer Engineering, Isfahan University of Technology, Isfahan, Iran

[b] Department of Bioelectrics and Biomedical Engineering, School of Advanced Technologies in Medicine, Medical Image and Signal Processing Research Center, Isfahan University of Medical Sciences, Isfahan 8174673461, Iran

[c] Deapartment of ophthalmology, Isfahan Eye Research Center, Isfahan University of Medical Sciences, Isfahan, Iran.

[*] **bahar.salafian@gmail.com**

[**] **m.pourazizi@yahoo.com**

## Abstract

The choroid is vascular tissue located underneath the retina and supplies oxygen to the outer retina; any damage to this tissue can be a precursor to retinal disease. Choroid is almost invisible in Enhanced Depth Imaging Optical Coherence Tomography (EDI-OCT) images and it is hard for an ophthalmologist to detect this layer manually. This paper presents an automatic method of choroidal segmentation from EDI-OCT images in neutrosophic space. In neutrosophic approach for image processing, extracting any information from an image and applying any process on it, is modeled by three sets including true, false, and indeterminacy sets. At first, we transform each image to the neutrosophic space; then, we calculate weights between each two nodes and apply the Dijkstra algorithm in order to detect Retinal Pigment Epithelium (RPE) layer. Based on RPE localization and applying gamma correction and homomorphic filter to false set, we segment choroid layer by similar approach used for RPE. The proposed algorithm is tested on 32 EDI OCT images of Heidelberg 3D OCT Spectralis from 11 people and is compared with manual segmentation. The results showed an unsigned error of 3.34 pixels (12.9 µm) for macular images and 6.55 pixels (25.3 µm) for prepapillary images; which macular error is lower than two other methods (7.71 pixels and 9.79 pixels). Futhermore, the proposed method on prepapillary data is novel and published works are concentrated on macular data. Identification of the boundary can help to determine the loss or change of choroid, which can be used as features for automatic determination of the retinal diseases.

## 1.Introduction

Choroid is a vascular layer lying between retina and sclera. This colorful layer contains a lot of capillaries to serve the supply of the iris and retinal light receptor cells [1]. In recent years, research has shown that the thickness of the choroid in some diseases including type-1 diabetes [2, 3], geographic atrophy [4], pathologic myopia [5], and retinopathy of prematurity [6-8] is reduced. Moreover, measuring thickness of this layer is useful for monitoring the progression of patients treated with VEGF[9, 10].

Optical coherence tomography (OCT) is a non-invasive imaging technique introduced by Huang et al. in 1991 [11]. Ophthalmologists utilize this method since they can take cross-sectional images of microscopic structures of living tissues.

In this method, a number of A-scans or linear scans create B-scan or cross-sectional image [11]. The light energy is used for imaging with the OCT device instead of the audio signal in ultrasound imaging, and the formation of the image depends on the optical properties of the structure of the tissues. Because of the high speed of light, direct measurement of the echo signal delay is not possible. As a result, OCT systems works based on the Low-coherence interferometry. OCT images consist a great amount of data; therefore, the non-automatic and visual analysis of such a large amount of information will be difficult for the ophthalmologist. The main purpose of the automatic segmentation is to help the ophthalmologist to diagnose and monitor eye diseases. Today, there are several types of OCT images including SD-OCTs, SS-OCTs, EDI-OCTs, among which, EDI-OCT images are commonly used for choroidial imaging due to more depth-of-field imaging [12]. Fig.1 shows an example of the choroidal layer location in one sample EDI- OCT image.

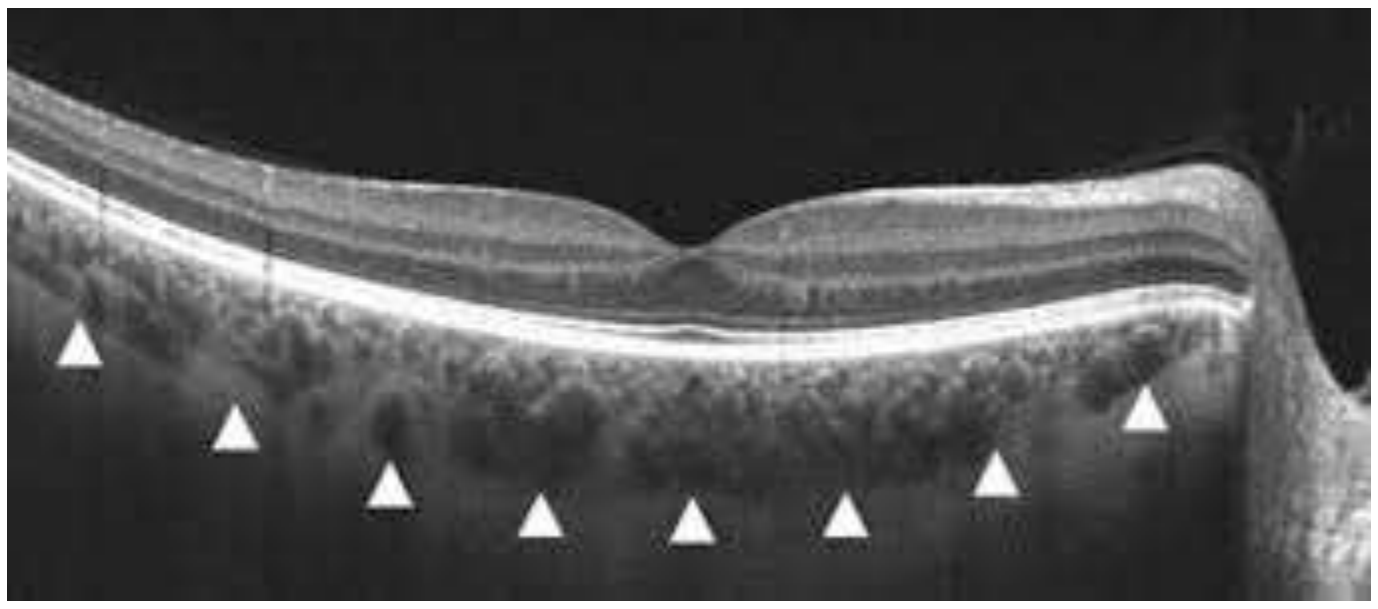

Fig.1.Location of choroid in a sample EDI-OCT image

## 2.Background and motivation

In recent years, a large number of automatic and semi-automatic methods have been reported in order to segment the choroid layer. table 1 summarizes a number of these algorithms.

Table.1. A summary of available automatic and semiautomatic methods for localization of choroidal layer

| Year | Author | Method | Data type | OCT acquisition region | Reported error value and method |
|---|---|---|---|---|---|
| 2012 | Li Zhang[13] | Segmented choroidal vessels through vesselness map, Multi-scale Hessian matrix and the tensor matrix | Zeiss Cirrus; Carl Zeiss Meditec, Inc., Dublin, CA(SD-OCT) | Macula | Average dice coefficient of the reproducibility 0.78±0.08 |
| 2013 | Hajar Danesh[12] | Multiresolution Texture Based Modeling in Graph Cuts | Heidelberg Spectralis HRA + OCT | Macula | Unsigned error 9.79 ± 3.29(pixel) |
| 2015 | Qiang Chen[14] | Segment RPE-BM-Choriocapillaris (RBC) and Choroid Scelera Interface (CSI) through thresholding, gradual intensity distance, graph min-cut-max-flow and the energy minimisation technique | The commercially spectral-domain device, Cirrus (HD-OCT ) | Macula | The mean Choroid Thickness (CTh) difference and overlap ratio are 6.72μm and 85.04%, respectively. |
| 2018 | Md AkterHussain[15] | Segment Internal Limiting Membrance (ILM) and RPE-BM-Choriocapillaris (RBC), approximate Choroid Scelera Interface (CSI) and Outer Choroidal Vessels (OCV),calculate edge weight | Heidelberg Spectralis HRA + OCT | macula | Mean root-mean-square error 7.71±6.29 (pixel) |

## 3.Proposed method

Fig.2 represents the block diagram of the proposed method. First of all, the gray image is transformed into the neutrosophic space (that converts the image into three sets of true, false and indeterminacy sets); then, graph shortest path method is employed in order to detect the retinal pigment epithelium (RPE) layer. After noise removal, the layers are flattened based on the RPE location; homomorphic filter and gamma-correction are employed to the false set to remove non-uniform intensities and to reduce light reflection of the RPE layer. Then, graph shortest path algorithm is used to detect the choroid location, the thickness map is achieved. Furthermore, the software provides the possibility of manual correction for users. Each step is elaborated in more detail in the following sections.

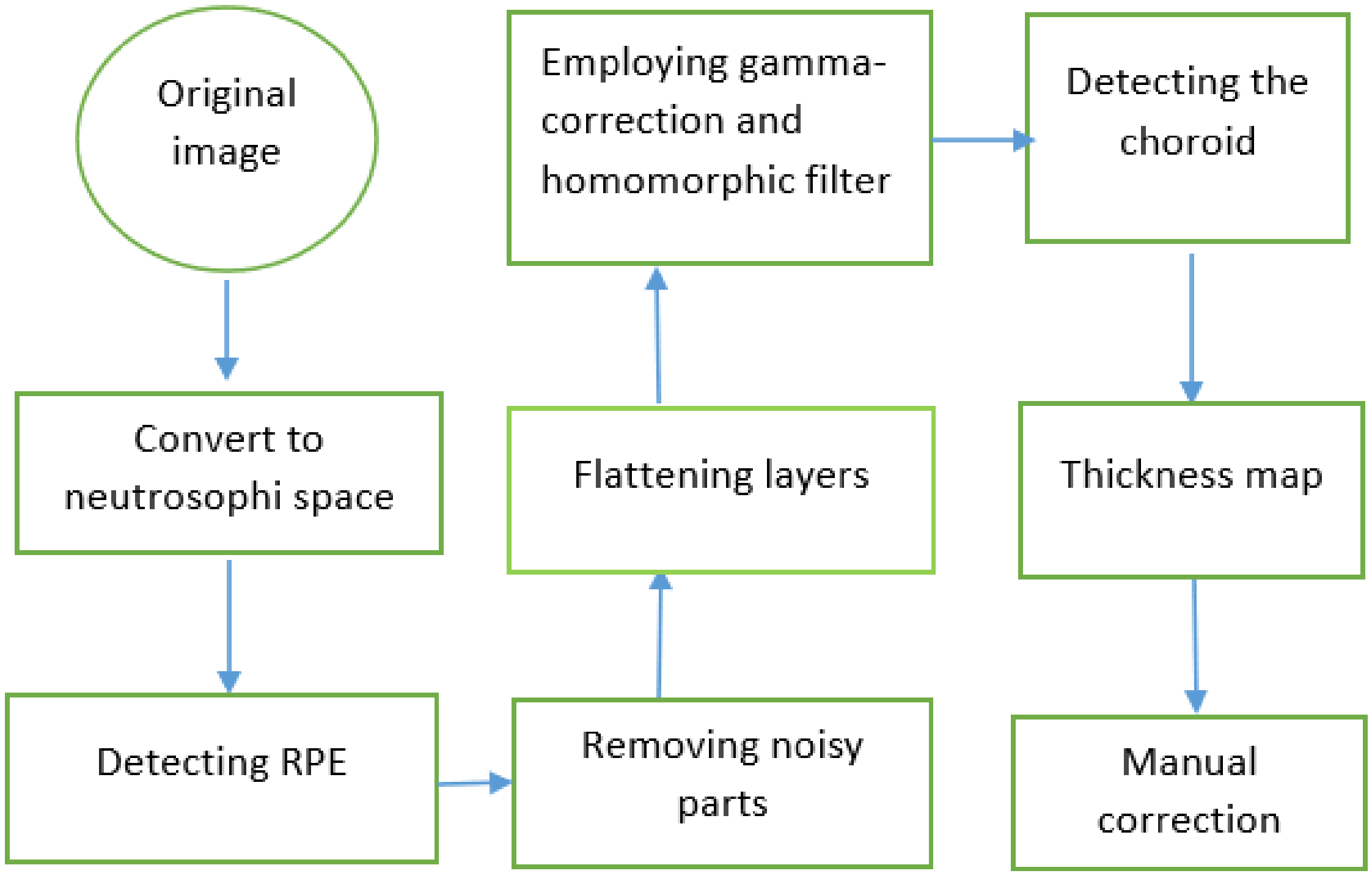

Fig.2. Flowgram of the proposed method

## 3.1. Neutrosophic theory

Neutrosophy is a branch of the philosophy which studies the nature and the scope of the neutralities and their interactions which is the basis of the neutrosophic logic and neutrosophic set. This theory was applied for image processing first by Guo et al [16] and has subsequently been used for other image processing applications including image segmentation [16-19], image thresholding [20], edge detection[21], and image clustering [22]. In neutrosophic approach for image processing, extracting any information from an image and applying any process on it, is modeled by three sets including true, false, and indeterminacy sets. In this approach, for each type of image (medical, satellite, natural, etc.), a specific set is defined for the indeterminacy of the pixels; therefore, any kind of noise in the image can be defined by introducing the appropriate set of indeterminacy [16].

### 3.2.1. Converting original image to neutrosophic space

An neutrosophic image is represented by $P_{NS}$ which contains F, T, and I sets.

Moreover, a pixel of the original image (g(i,j)) which is shown by
in the neutrosophic space. The formulas showing the $P_{NS}(i,j) = \{T(i,j), I(i,j), F(i,j)\}$
relationship between pixels in image scope and neutrosophic space are represented in (1-5).

$$T(i,j) = \frac{\bar{g}(i,j) - \bar{g}_{min}}{\bar{g}_{max} - \bar{g}_{min}} \quad (1)$$

Where:

$$\bar{g}(i,j) = \frac{1}{\omega^2} \sum_{m=-\omega/2}^{\omega/2} \sum_{n=-\omega/2}^{\omega/2} g(i+m, j+n) \quad (2)$$

$$F(i,j) = 1 - T(i,j) \quad (3)$$

$$I(i,j) = \frac{\delta(i,j) - \delta_{min}}{\delta_{max} - \delta_{min}} \quad (4)$$

Where:

$$\delta(i,j) = |g(i,j) - \bar{g}(i,j)| \quad (5)$$

Where ,the $g(i,j)$ is the intensity of the $(i,j)$ , $\bar{g}(i,j)$ is local mean of the $g(i,j)$ , and is the absolute difference amount between these two values. Min and max $\delta(i,j)$ appendix represent the smallest and largest value in the matrix. In the above relationships, an average filter, with window size of 1×5, is employed in order to get the $\bar{g}$ image.

Assuming that our purpose is dividing a gray image into two parts (object and background) with object describing bright regions. T set represents the probability that each pixel belongs to the object, and I set displays probability of the uncertainty; for instance, for a noisy pixel, the more the value of $\delta(i,j) - \delta_{min}$ is, the more indeterminacy (I) is. In addition, the F set represents probability of belonging to background pixels and is obtained by $1-T$ . Fig.3 displays an example of this transformation.

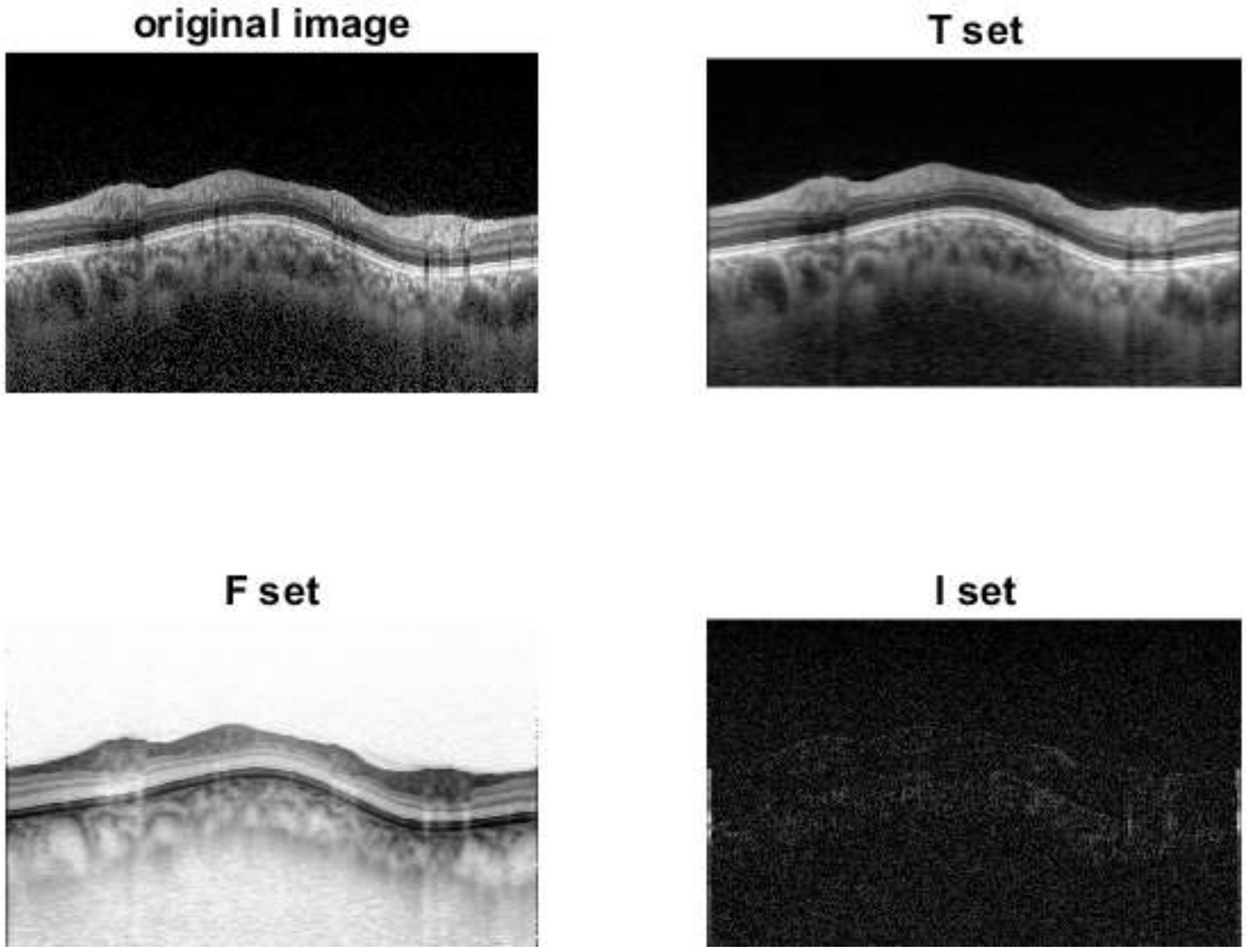

Fig.3. An example of neutrosophic conversion

### 3.2.2. Entropy of the image in neutrosophic space

In gray images, entropy is used as a criterion for distribution of the pixels' intensities; as a result, the pixels' of an image with non-uniform intensity will have minimum entropy. The entropy of a neutrosophic image is the sum of the entropies of the T, I, and F sets.

$$En_{NS} = En_T + En_I + En_F \tag{6}$$

Where:

Entropy of T: $En_T = -\sum_{min(T)}^{max(T)} P_T(i) ln P_T(i)$ (7)

Entropy of I: $En_I = -\sum_{min(I)}^{max(I)} P_I(i) ln P_I(i)$ (8)

Entropy of F : $En_F = -\sum_{min(F)}^{max(F)} P_F(i) ln P_F(i)$ (9)

Where $P_T(i)$ represents probability that i belongs to the T set.

### 3.2.3. α-mean operation

In order to make connection between F and T with I, it is necessary to make changes in the T and F sets that affect the distribution of the elements and entropy of the I set. Hence, α-mean operation is employed based on the following formulas [16]:

$$\bar{P}_{NS}(\alpha) = P(T_\alpha, I_\alpha, F_\alpha) \quad (10)$$

$$T_\alpha(i,j)=\begin{cases}T(i,j) & if I(i,j)<\alpha \\ \bar{T}(i,j) & else\end{cases} \quad (11)$$

$$\bar{T}(i,j)=\frac{1}{\omega^2}\sum_{m=-\omega/2}^{\omega/2}\sum_{n=-\omega/2}^{\omega/2}T(i+m,j+n) \quad (12)$$

Where ω is equal to 5.

$$F_\alpha(i,j)=\begin{cases}F(i,j) & if I(i,j)<\alpha \\ \bar{F}(i,j) & else\end{cases} \quad (13)$$

$$\bar{F}(i,j)=\frac{1}{\omega^2}\sum_{m=-\omega/2}^{\omega/2}\sum_{n=-\omega/2}^{\omega/2}F(i+m,j+n) \quad (14)$$

Furthermore we have:

$$\bar{\bar{T}}(i,j)=\frac{1}{\omega^2}\sum_{m=-\omega/2}^{\omega/2}\sum_{n=-\omega/2}^{\omega/2}\bar{T}(i+m,j+n) \quad (15)$$

$$\bar{\delta}_T(i,j)=\left|\bar{T}(i,j)-\bar{\bar{T}}(i,j)\right| \quad (16)$$

And:

$$I_{\alpha}(i,j) = \frac{\bar{\delta}_T(i,j) - \bar{\delta}_{Tmin}}{\bar{\delta}_{Tmax} - \bar{\delta}_{Tmin}} \quad (17)$$

### 3.3. Graph theory

Graph theory is used in various image processing applications, including image segmentation. The image segmentation methods based on the graph theory are divided into four categories such as graph cut, graph shortest path, minimum spanning tree, and Markov random model [23]. In this study, we apply the graph shortest path for detecting the RPE and choroid layers. Consider the graph G= (V, E) graph, in which V is the set of vertices and E is the set of edges and each edge has the weight of $w_e$. Finding the shortest path between each two pixel leads to a path that can be considered as the boundary between image objects. So one can split the image by creating different borders [23].

#### 3.3.1. RPE layer segmentation

The RPE layer segmentation plays a crucial roles and will increase the speed of the process because we can limit the ROI in order to detect the choroid layer. For this purpose, we pass the T set through a filter in order to achieve vertical gradient of the image. Fig.4 demostrates an example of this output.

$$VerGrad = T * H, \qquad H = \begin{bmatrix} 2 \\ 0 \\ -2 \end{bmatrix} \quad (18)$$

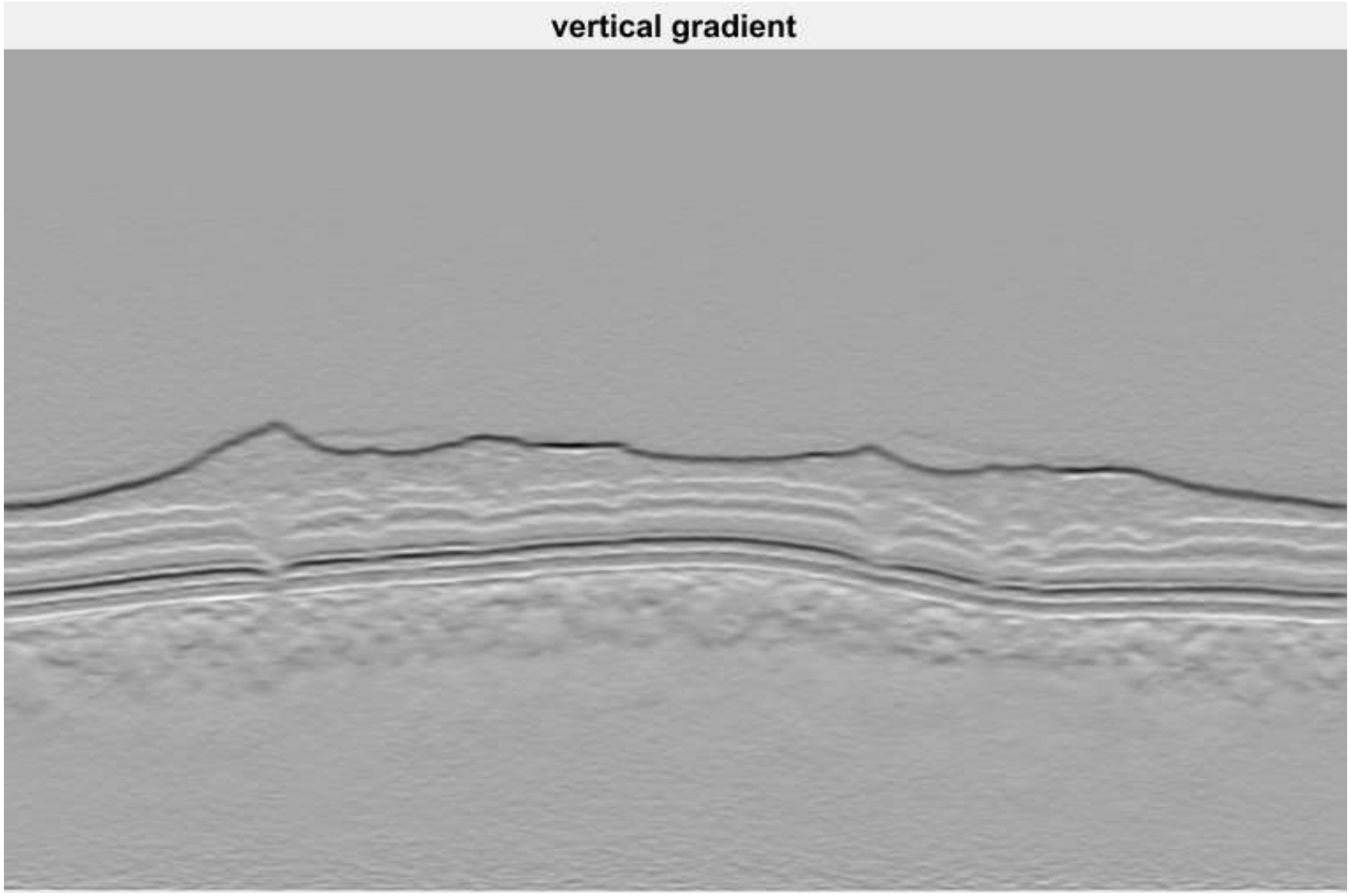

Fig.4.Vertical gradient of an image

We construct a graph for each OCT image through mapping each pixel to single node of a graph. We just consider the local relationship between pixels. Therefore, by making connection among each pixel and its 8 neighbors, the 8-regular graph will be created. Formula (19) represents calculating the weight between two desired pixels.

$$W((a1,b1),(a2,b2)) = 4\times MaxG - VerGrad(a1,b1) - VerGrad(a2,b2) - mean(D) \qquad (19)$$

Where, W is the weight between the nodes, MaxG illustrates the maximum gray level of the image, $mean(D)$ is the mean value of a set of D pixels which are lying above the RPE, here D is equal to 10 pixels, and $VerGrad$ is the vertical gradient of that pixel.

Because of the H filter, maximum vertical gradient of each pixel is two times greater than the MaxG. As a result, the maximum value of *VerGrad(a1,b1)+VerGrad(a2,b2)* will be four times greater than *MaxG*. In fact, the pixels of the RPE layer have the maximum vertical gradient since they are locating between the dark (choroid vessels) and the light (photoreceptor layer) regions. Moreover, the lighter layer is lying above the darker layer. With this definition of the graph, RPE pixels will have the minimum weight and will be chosen by the shortest path algorithm. Because pixels of other layer above the RPE may have the similar situation (maximum vertical gradient), we subtract the *mean(R)* form the formula, in which R is the upper pixels of the RPE. The value of this mean is huge because the intensities of these pixels are high which lead to the minimum weight for the RPE boundary [24]. In order to implement this algorithm in MATLAB, at first, we create an adjacency matrix that its dimension is [MN×C] in which C are neighboring pixels (which is 8 in this application) and MN is production of multiplying rows and columns of the image; thus, the adjacency matrix has M×N×C elements. Fig.5 demonstrates one sample of RPE detection by this approach.

### 3.3.2. Dijkstra algorithm

After calculating weight between each two nodes, dijkstra algorithm is employed in order to specify the minimum weighted path [25]. Then, in order to lead the algorithm toward detection of the RPE layer, we add a small value to the first and last column of the image as a minimum weight of the layer. It should be noted that this value must be lower than all calculated weights. In this research, is equal to $10^{-5}$ [26]. Fig.5 demonstrates the role of minimum columns in $W_{min}$ correct localization of the RPE layer. Finally, the RPE layer can be identified correctly as shown in fig.6.

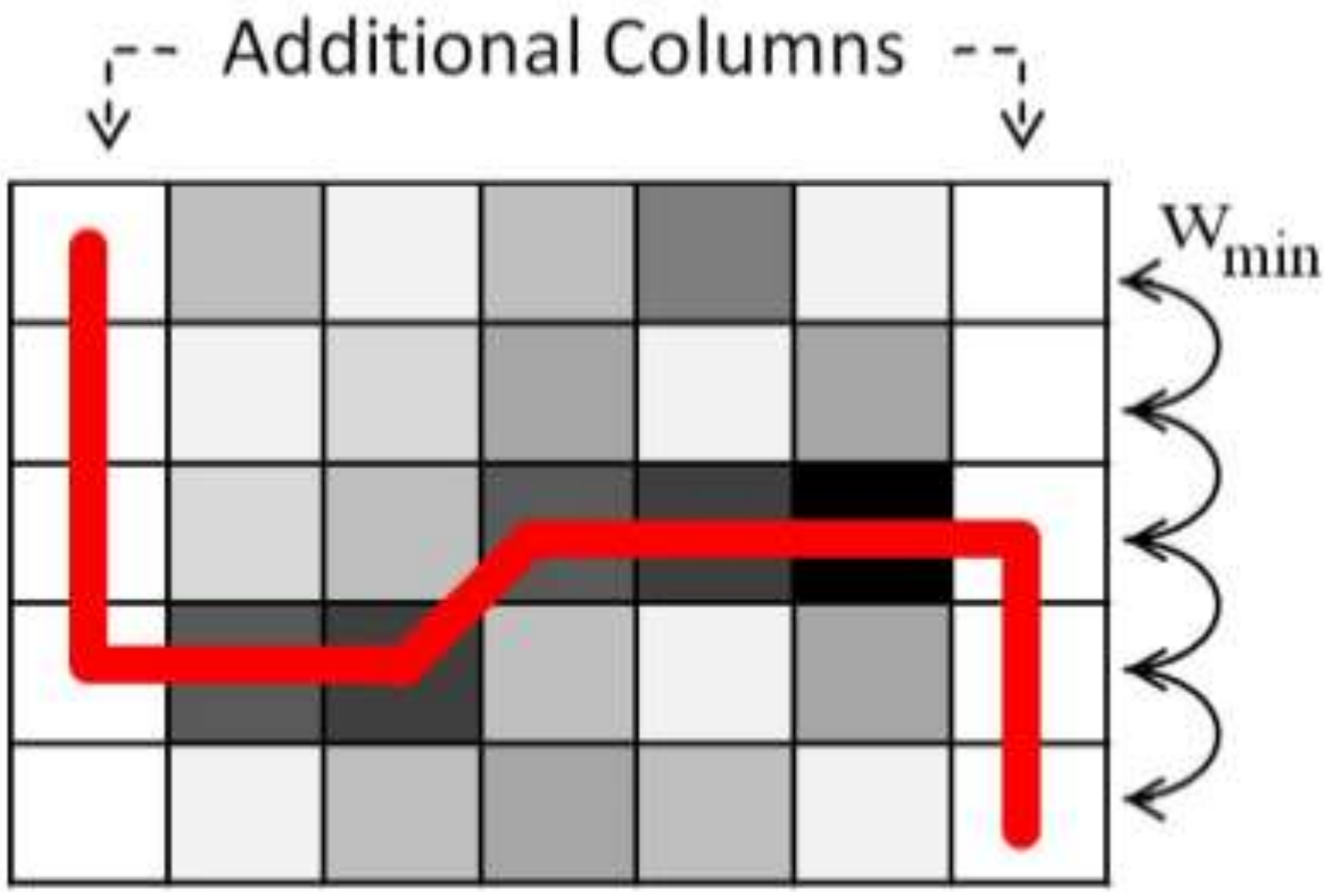

Fig.5. auto-setting the initial point

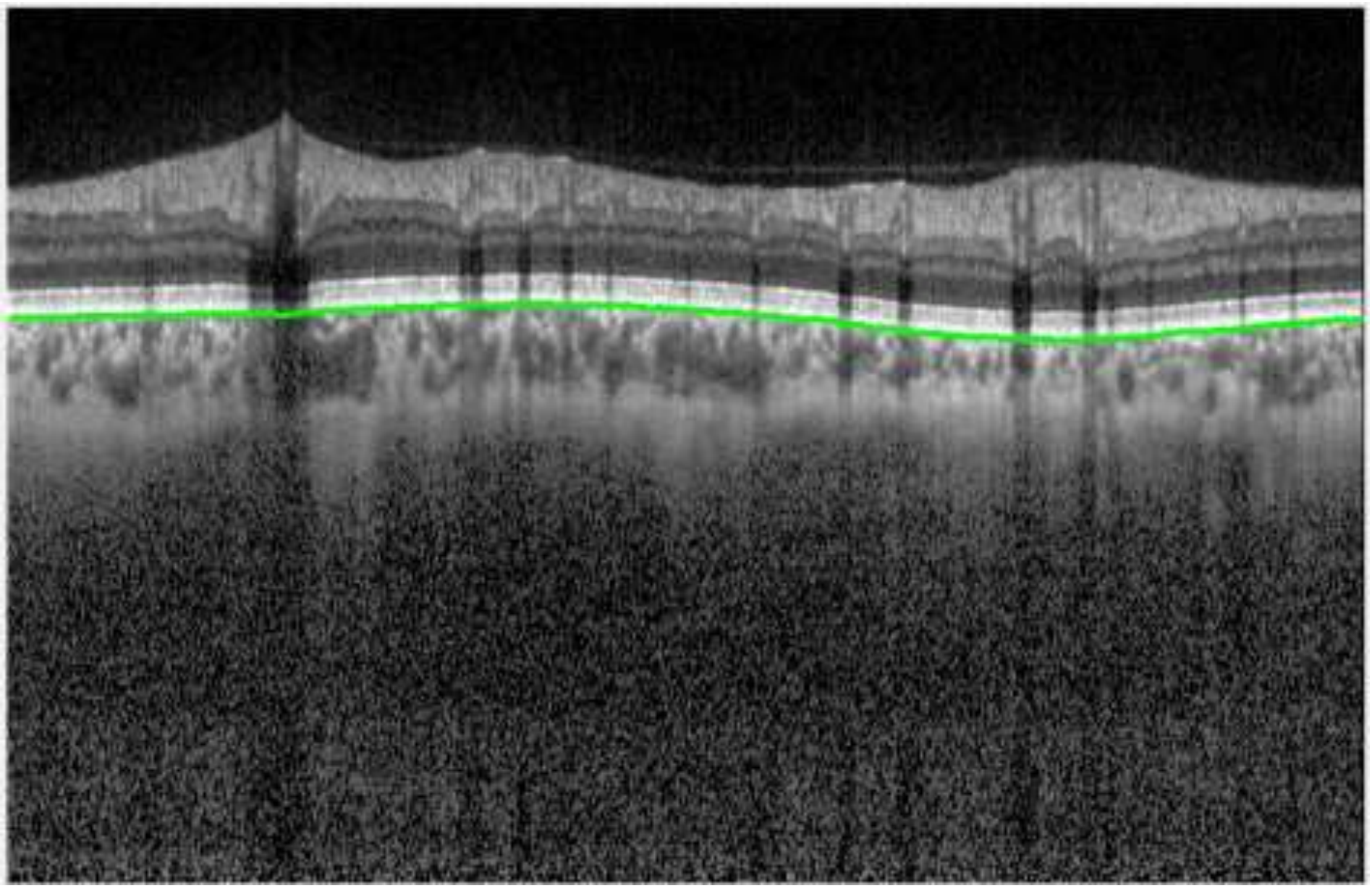

Fig.6. RPE detection of the proposed method (green line represents RPE layer)

## 3.4. Image flattening

As mentioned in Section 3.3.1, a graph is segmented by computing the minimum weighted path from a short node to an end node. Inherent in this method is the shortes geometric path to be found, since fewer traversed nodes results in a lower total weight. As a result, features with strong curvature or other irregularities, even with relatively strong gradient, are disadvantaged since their paths do not reflect the shortest geometric distance [26]. For solving this problem, we flatten layers based on the RPE location and limit the ROI from RPE to the bottom of the image [27]. To smooth layers, first we find the lowest part of the RPE boundary; then, we calculate the difference between other pixels of the RPE boundary and that lowest part. In the next step, corresponding to the difference of each column, we move it toward the bottom of the image and add the removing part at the beginning of the column (noise data). Fig.7 indicates this translation.

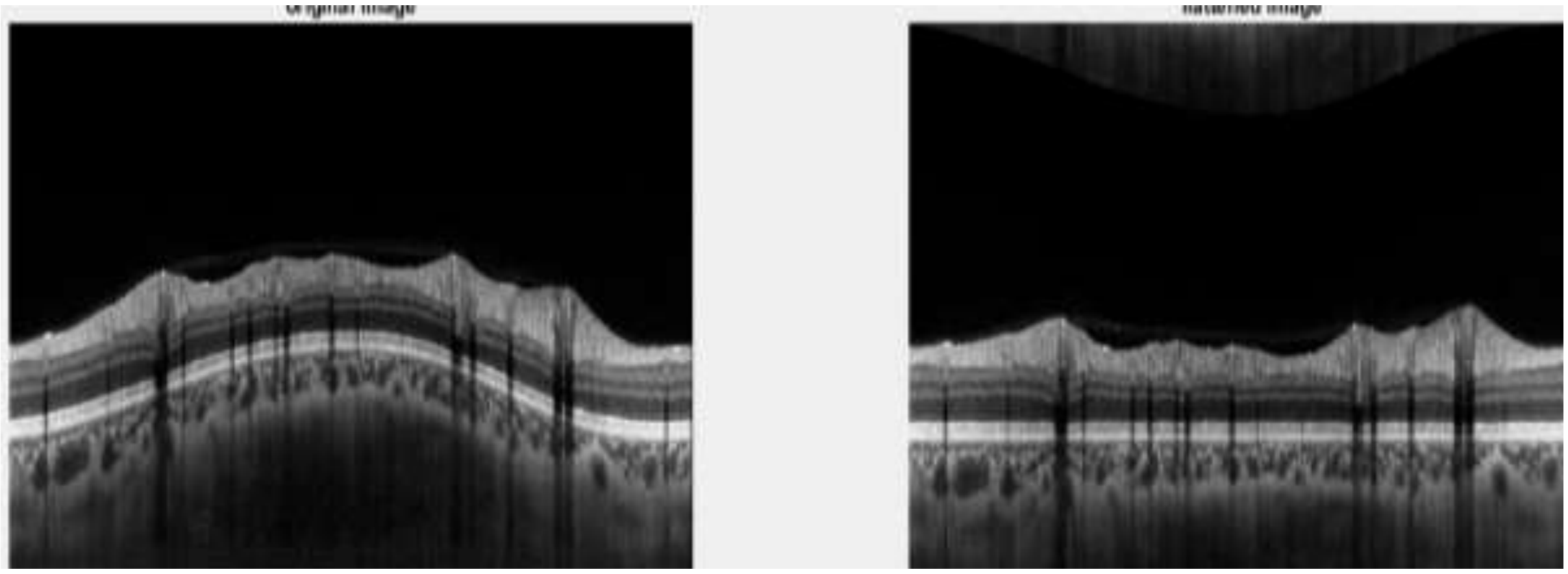

Fig.7. Right side image is flattening image of the left side image

## 3.5. Choroid layer segmentation

For segmenting the choroid layer, we select the F set rather than T set since the reflection of the RPE intensitis are less visible in F set; therefore, we employ the graph shortest path algorithm to this image.

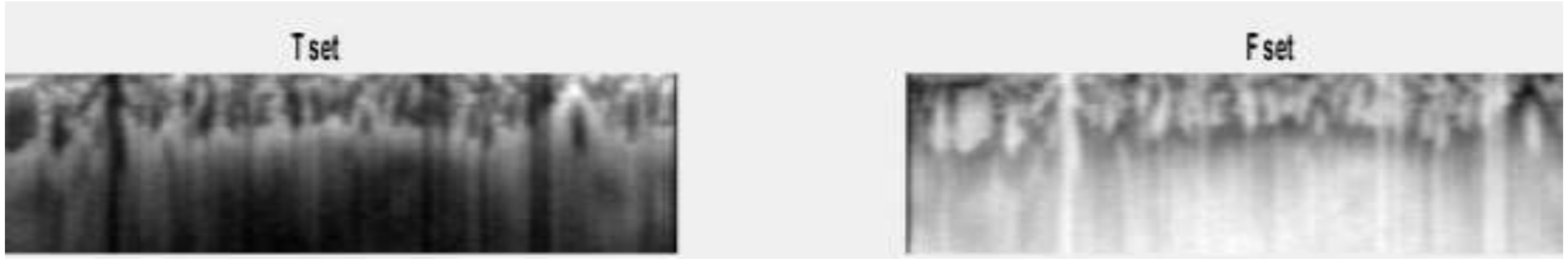

Fig.8. F set and T set of the ROI

### 3.5.1. Gamma correction

As it is shown in fig.10, the intensities of the choroid vessels (green circles) are more than the choroid layer (yellow arrows). As a result, vertical gradient will be higher in vessels and the algorithm will detect the vessels rather than choroid. As a solution, we employ the gamma correction (eq.20) which leads to brightening dark areas of the image while maintaining the intensity of light area. The result is shown in fig.11.

$$F_{out} = 255^{1-\gamma} \times F^{\gamma} \quad (20)$$

Where γ is equal to 0.2.

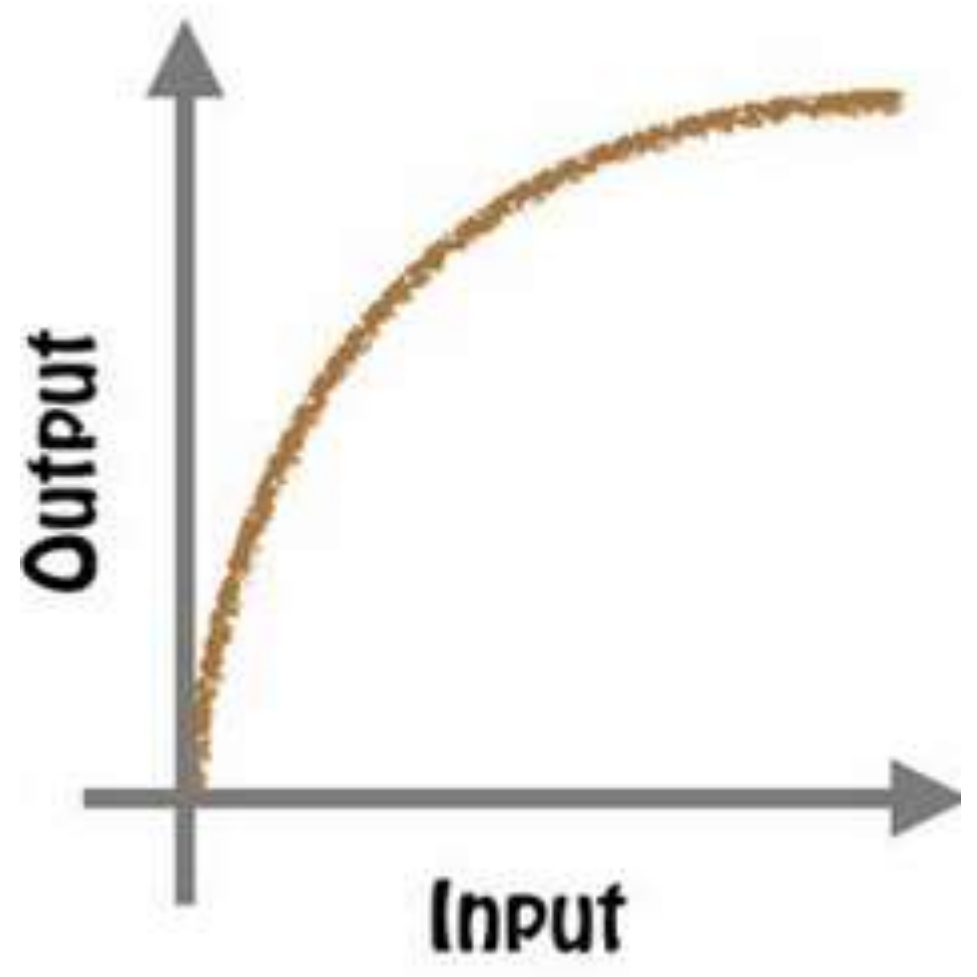

Fig.9. the diagram of the gamma correction for 0<γ<1

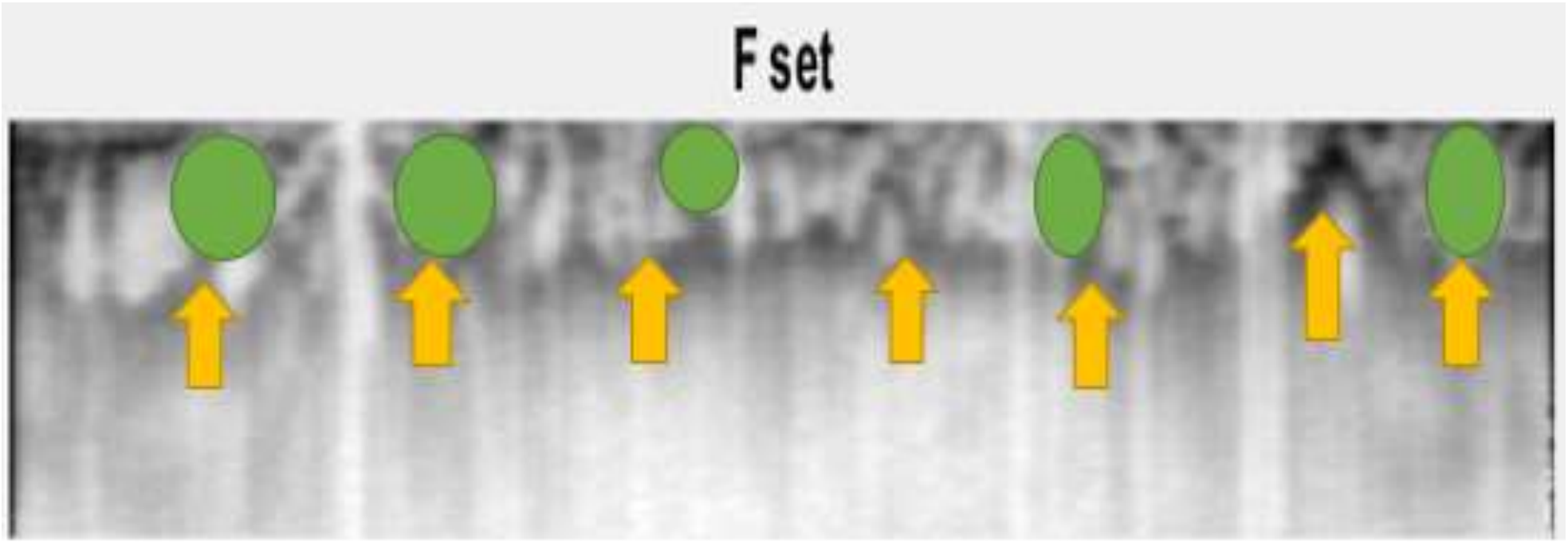

Fig.10. The F set of ROI before gamma correction (yellow arrows represent choroid boundary and green circles illustrate choroidal vessels)

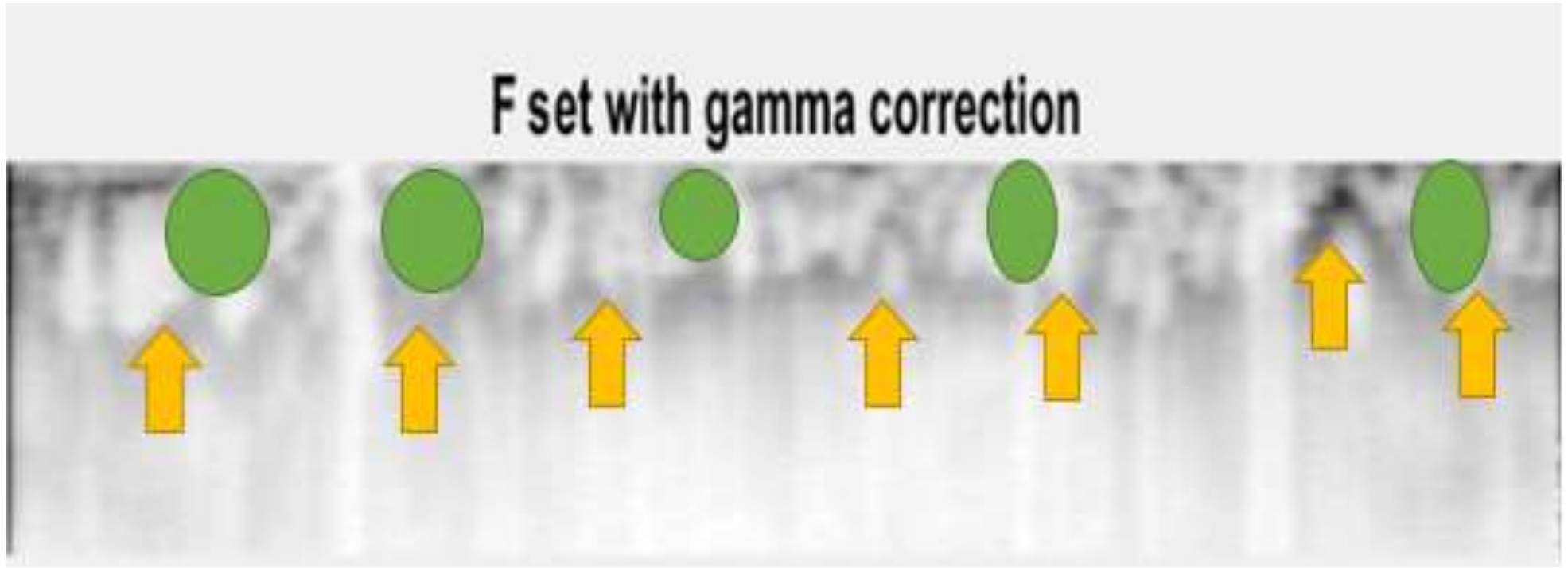

Fig.11. The F set of ROI after gamma correction (yellow arrows represent choroid boundary and green circles illustrate choroidal vessels)

### 3.5.2. Homomorphic filter

As demonstrated in fig.11, non-uniform illuminations are available in resultant image. This kind low frequency artifact is located in the low frequency part of the image. We reduce this artifact by attenuating the DC part of the image via homomorphic filter. In homomorphic formula, three parameters play the crucial role, $\sigma, \gamma_H, \gamma_L$ that are variance

, the gain of the high frequency, and the gain of the low frequency respectively. The values are selected empirically and set to 3.2, 1, 0, respectively.

$$H(u,v)=(\gamma_H-\gamma_L)(1-exp^{-\frac{D^2(u,v)}{2\sigma^2}})+\gamma_L; \quad D^2(u,v)=(u-M/2)+(v-N/2) \qquad (21)$$

Where, $(u,v)$ is the coordinate in frequency domain, and $(M,N)$ is dimension of the original image. Then, we calculate the logarithm of fourier transformation from the image and reorganise the filter and image to the center of coordinate to correct application of the filter to the image. Fig.12 and 13 represent the output of the algorithm without and with the homomorphic filter, respectively.

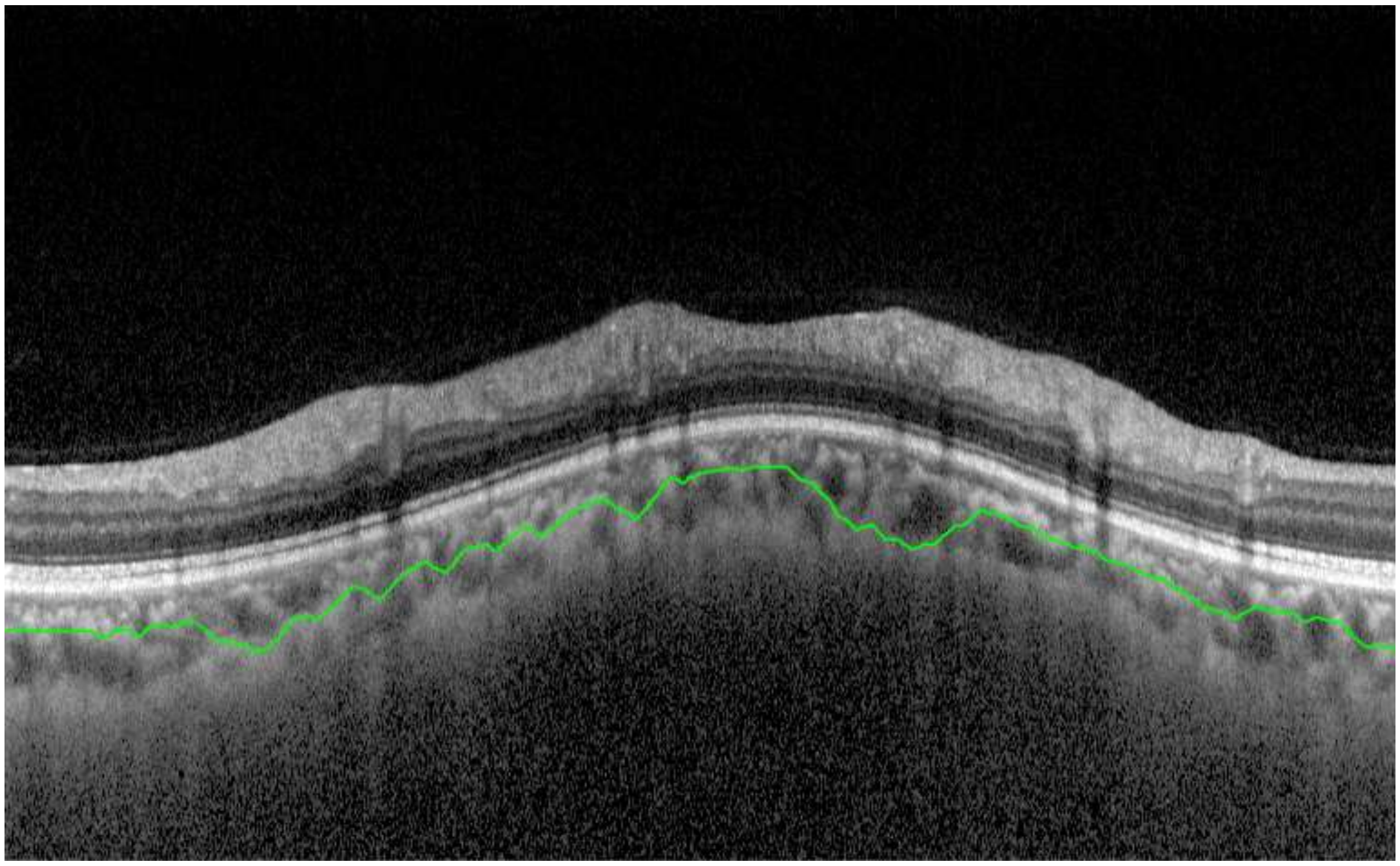

Fig.12. An output of the algorithm without homomorphic filter

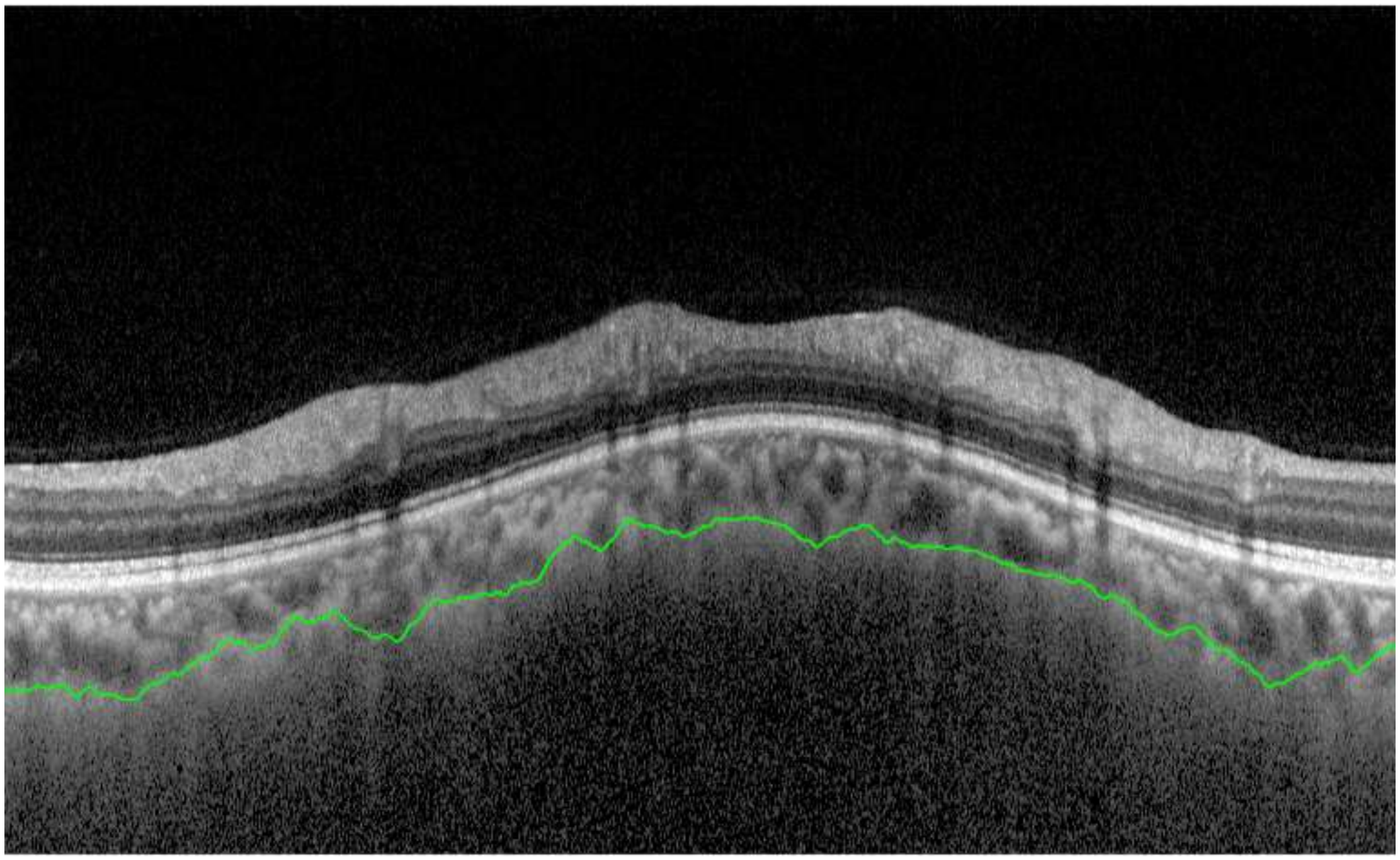

Fig.13. An output of the algorithm without homomorphic filter

### 3.5.3. segmentation of Choroid Layer

In final step, the negative gradient is used to calculate the weight of the nodes. This section is reasonable since the boundary of the choroid is lying above the dark and below of the light region. Then, the graph shortest path algorithm is applied to detect this layer. Fig.14 represents an example of the segmentation with proposed method.

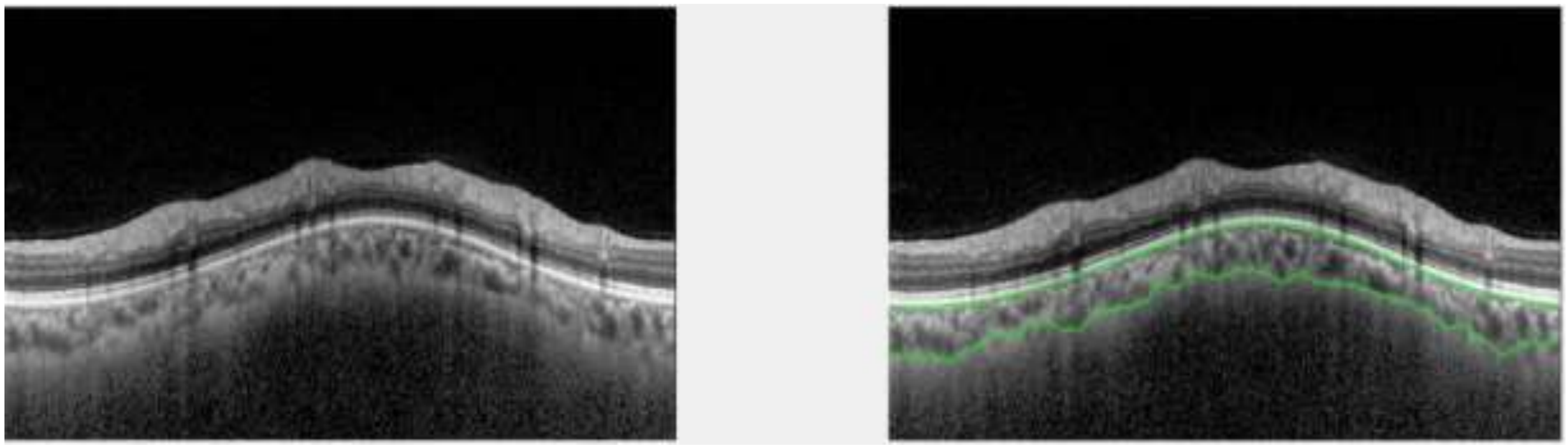

Fig.14. choriod segmentation based on the proposed method (the lowest green line is choroid location)

### 3.6. Thickness map

For achieving this thickness images, and to show them in colormap similar to maps produced for other retinal layers by OCT devices, we determine the location of both RPE and choroid layers on macular images; then we subtract them from each other, convert pixels to millimeter with multiplying each pixel by 0.00387167 (resolution), and show the color images. Now, the ophthalmologist can compare the color of each area in order to find the degree of thinning in choroid. Fig.16 indicates a sample of these images.

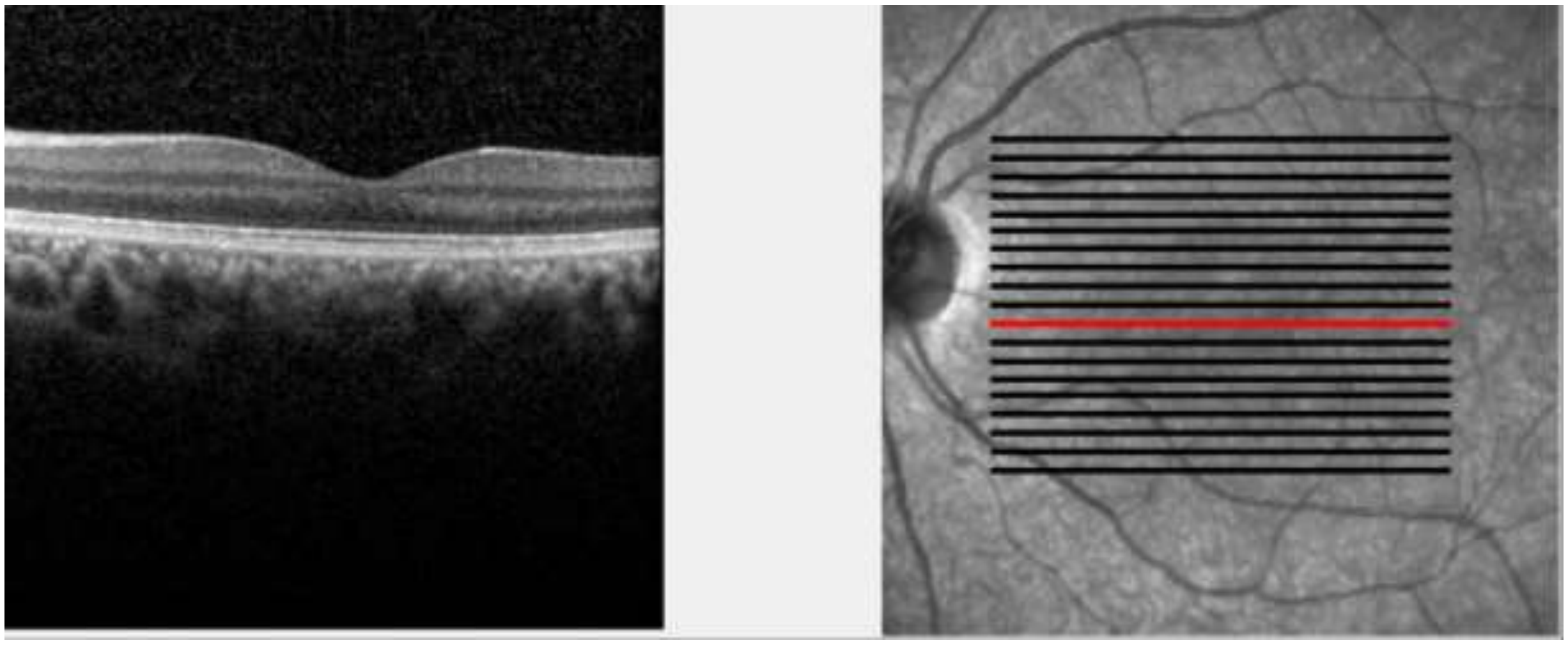

Fig.15. A sample of the macular images

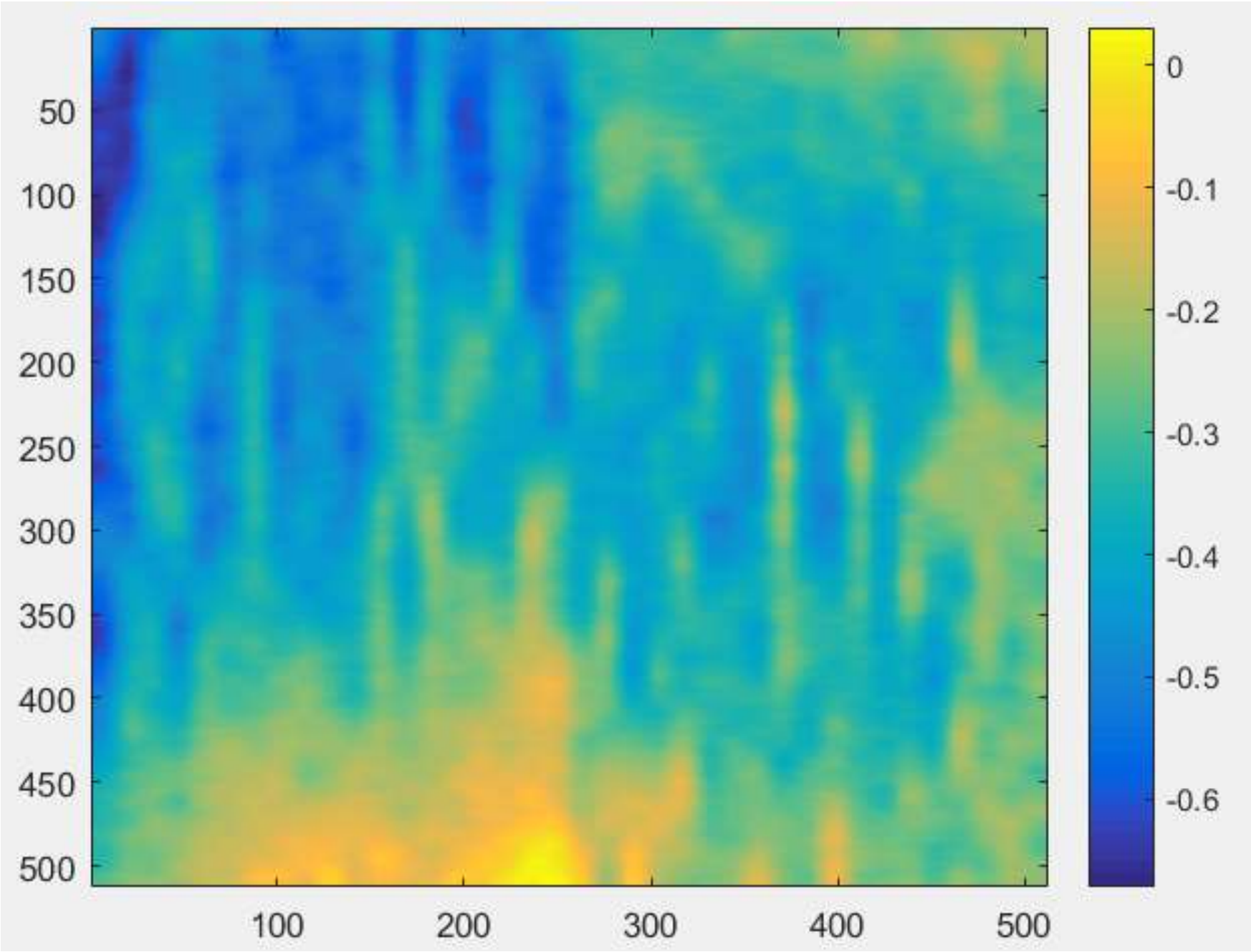

Fig.16. An output of the thickness map image

## 3.7. Manual correction

After automatic detection of the choroid layer, the software provides the possibility of needed manual correction for the user. In manual correction, as the usre selects two points from in region, like A and B points in fig.18. The automatic location lying between these two nodes is then removed, and vertical lines are drawn between two sides to get desired location. Finally, new location of the choroid will be visible after interpolation.

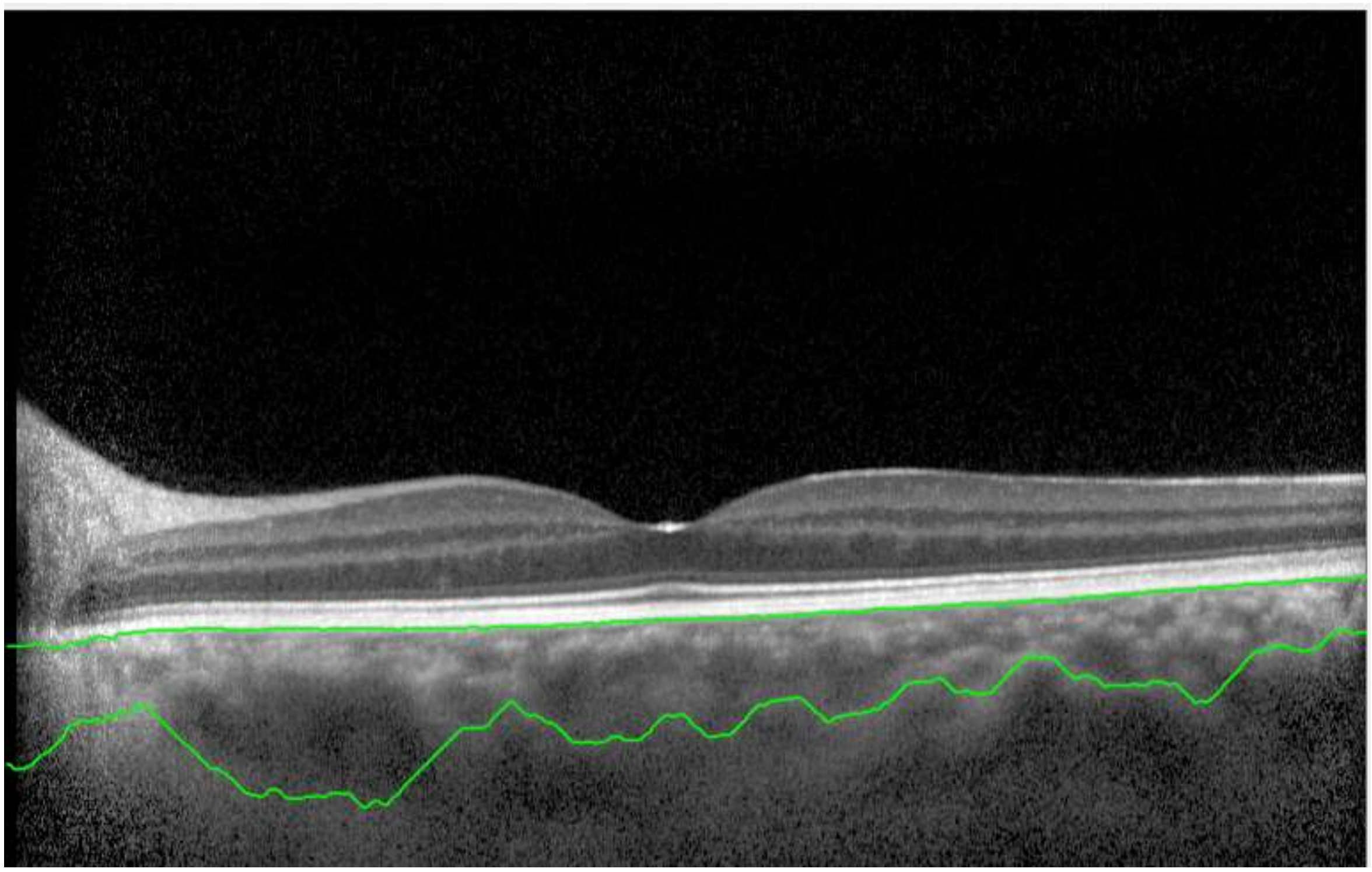

Fig.17. Choroid segmentation of the proposed method

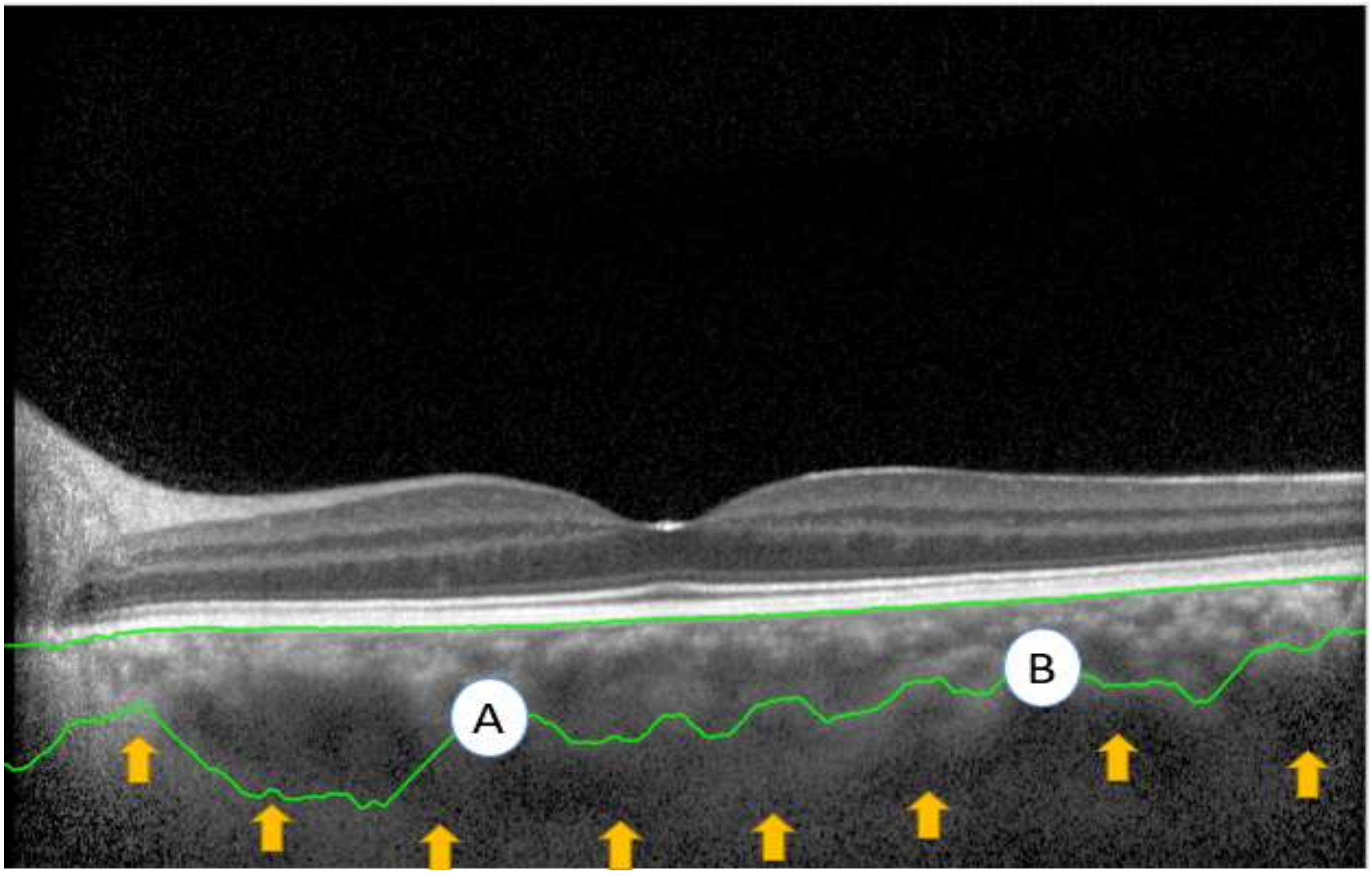

Fig.18. The green line is choroid location of the automatic method, yellow arrows are the desired place of the choroid for the user, and A and B points are selected by user

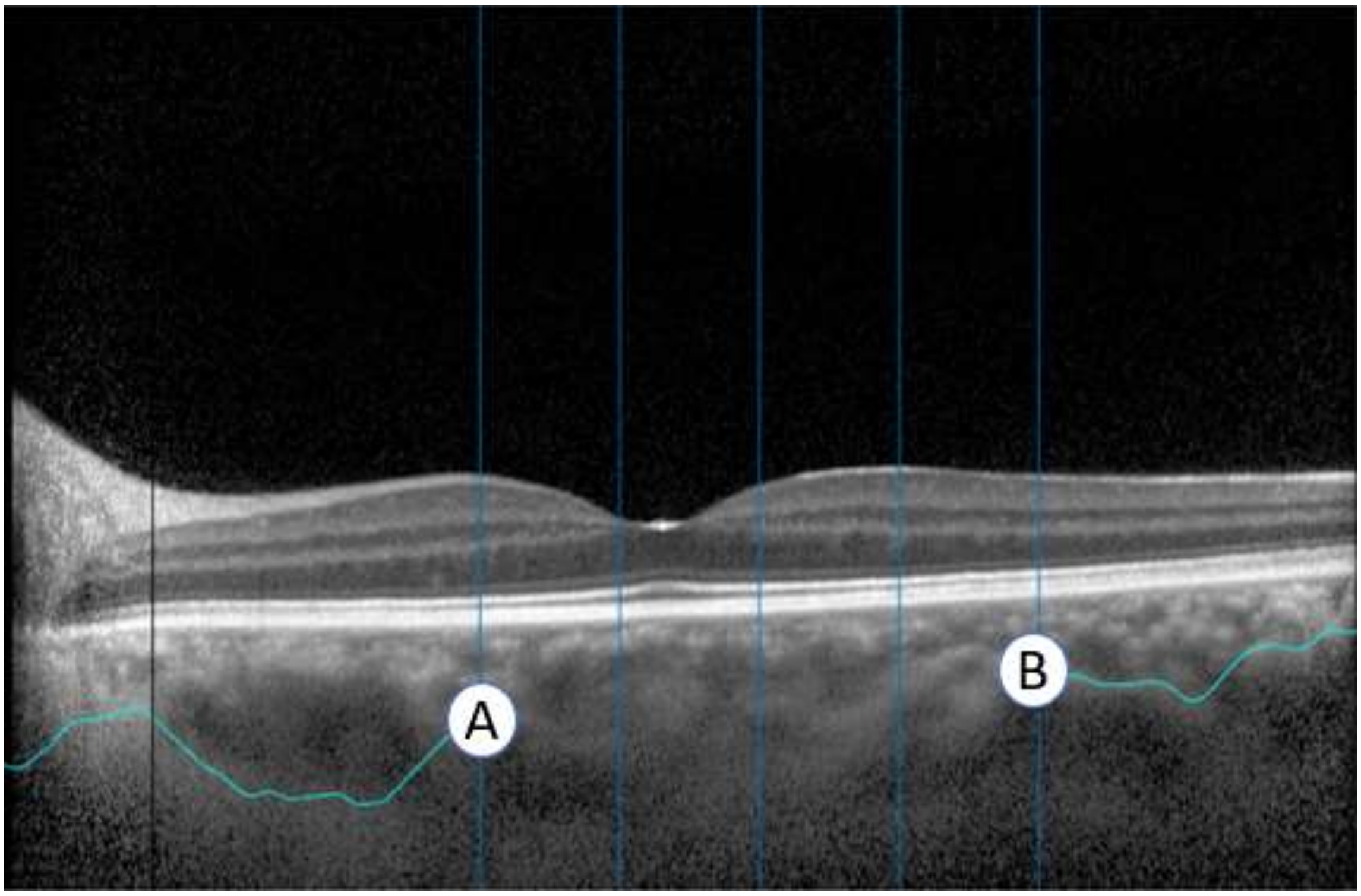

Fig.19. removing the choroid between selected points

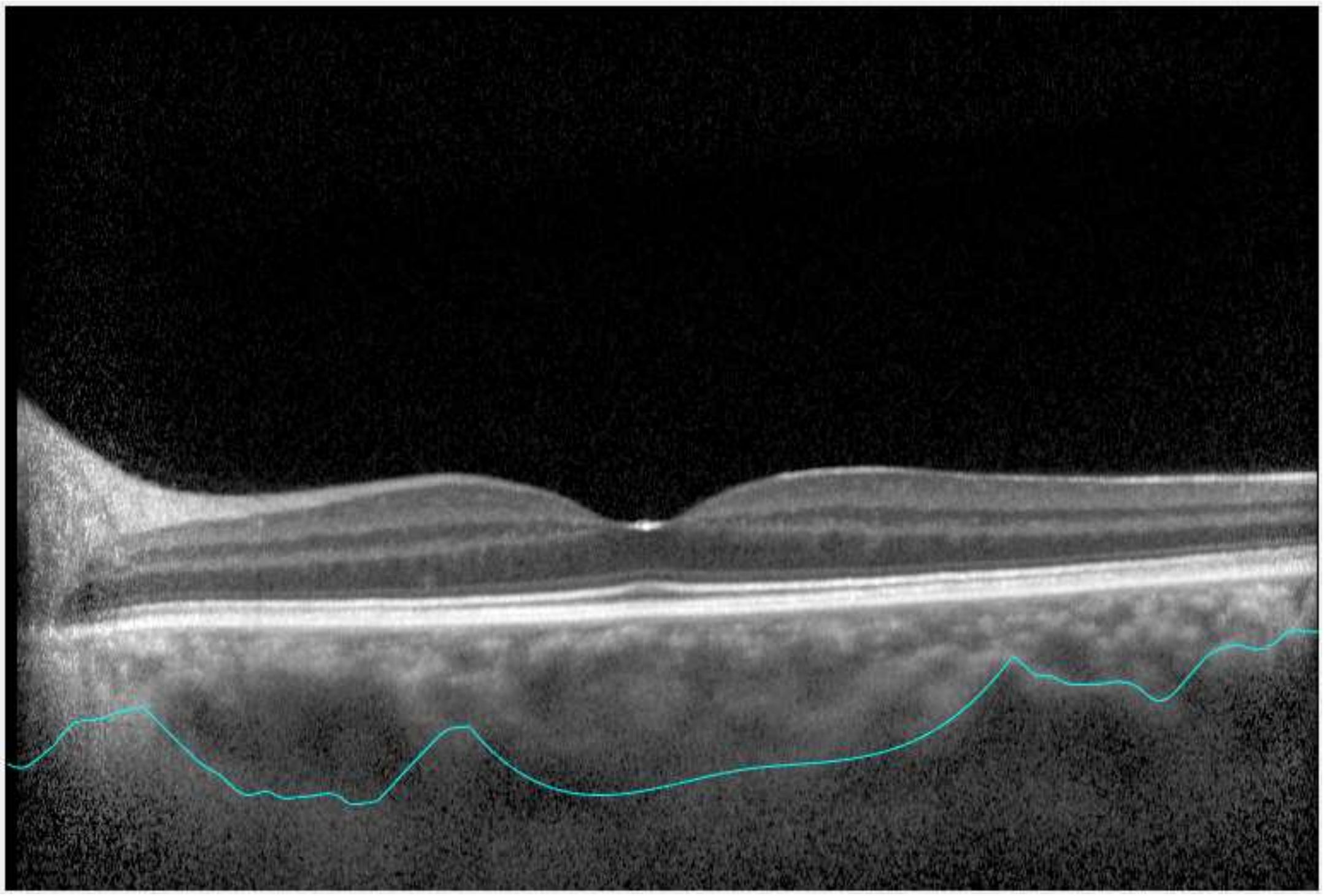

Fig.20. Blue line represents the output of the manual correction

## 4. Experimental setup

Our experimental setup consisted 32 EDI-OCT images of 11 people were examined, 17 numbers of images are circular and 15 numbers of them are macular. Circular or peripapillary images are images taken from the ONH, while macular images are ones taken around the macula. Figures 21 and 22 illustrate this subject. All people are imaged using EDI-OCT imaging on Heidelberg Spectralis HRA+OCT at the the Ophthalmology clinic of Faiz Hospital in Isfahan under the supervision of Dr. Alamzadeh. In addition, 7 of the people have MS and 4 of them are healthy. It should be noted that The mean subfoveal choroidal thickness is reduced significantly in MS patients [28].In this study, our data has two different dimension, 496×768 and 496×512 which their resolution is 3.87167 μm.

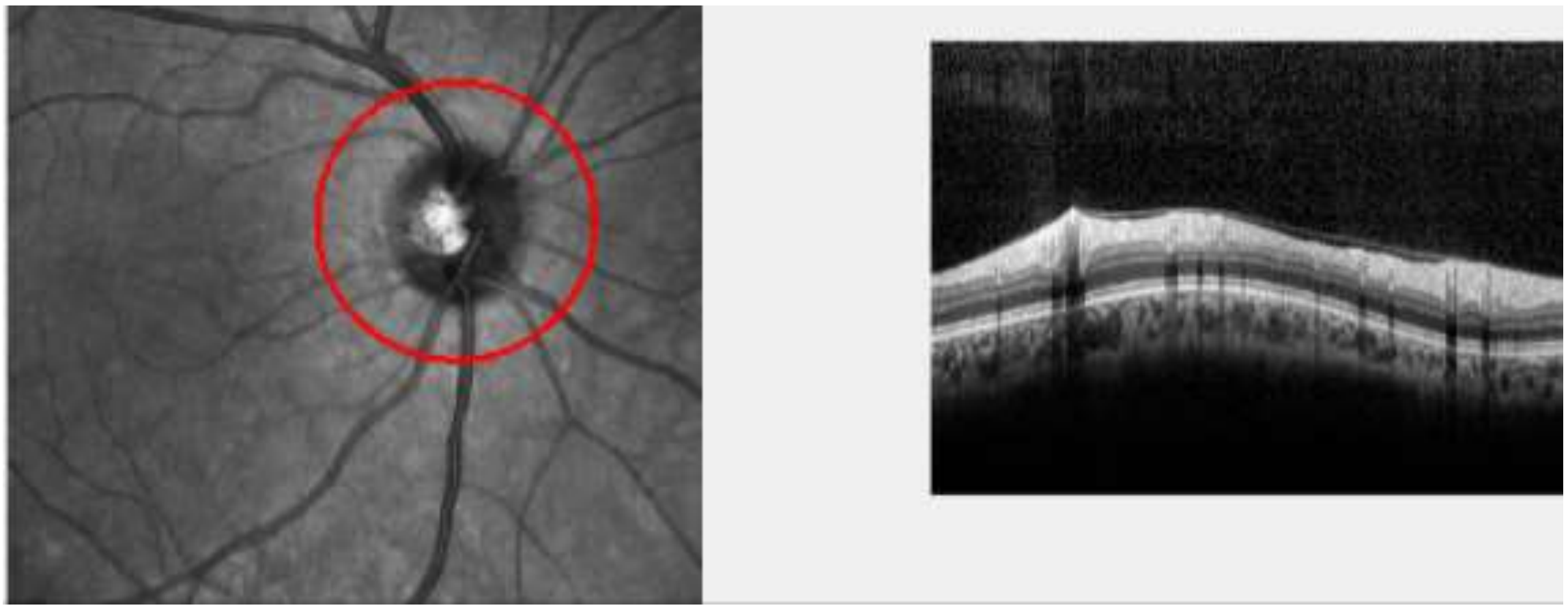

Fig.21. a sample of the peripapillary images

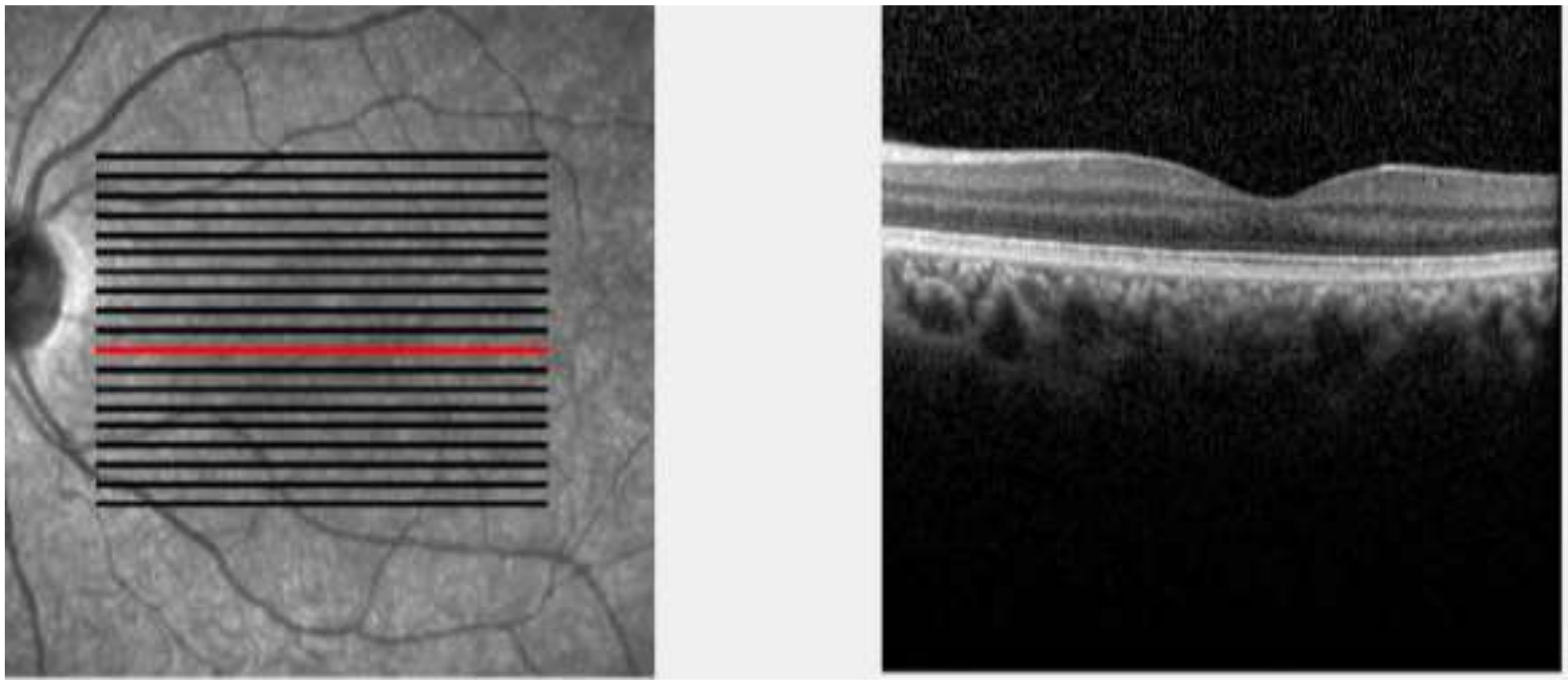

Fig.22. A sample of macular images

## 5. Results

The implementation of the proposed method takes about 60 seconds on MATLAB and Statistics Toolbox Release 2016b, The MathWorks, Inc., Natick, Massachusetts, United States, on Laptap with Microsoft Windows 10 x64 edition, Intel core i5 Duo CPU at 2.8 GHz, 8 GB RAM.To provide numerical explanation of the proposed method, the localization error is calculated in this section. We subtract the values labeled by the ophthalmologist out of the output of the algorithm. Fig.23 is an example of a labeled images.

Our results show better accuracy compared to the automatic method of M.A. Hussain et al.[15] and Danesh et al.[12] as presented in Table.2. Since non of the author does not release the data, we apply numerical comparison in this study. According to the table, output error for [15] and [12] are about 7.71 and 9.71 pixels respectively. However, this value is 3 pixels in poroposed method and shows better result for this issue. Fig.24 and fig.25 represent the outputs of the proposed method and the labeled images in both prepapilary and macular images (the red circles refer to the location determined by the ophthalmologist and the green line is the output of the proposed method). The error values of the unsigned errors are shown in the table.2. The ophthalmologist determines the limited points on each image; as a result, we cannot represent our error in terms of area or overlap.

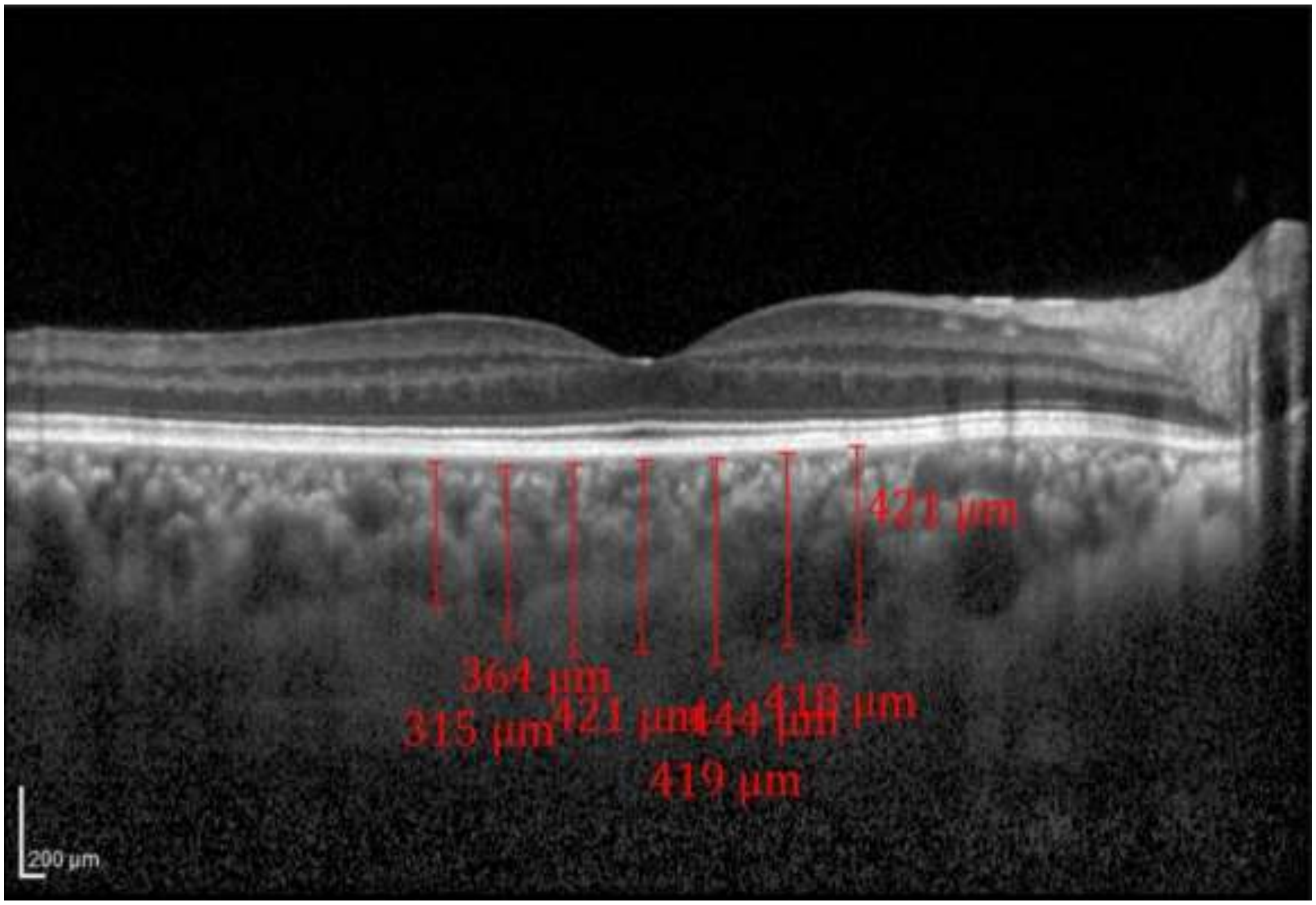

Fig. 23. A labeled image by the ophthalmologist

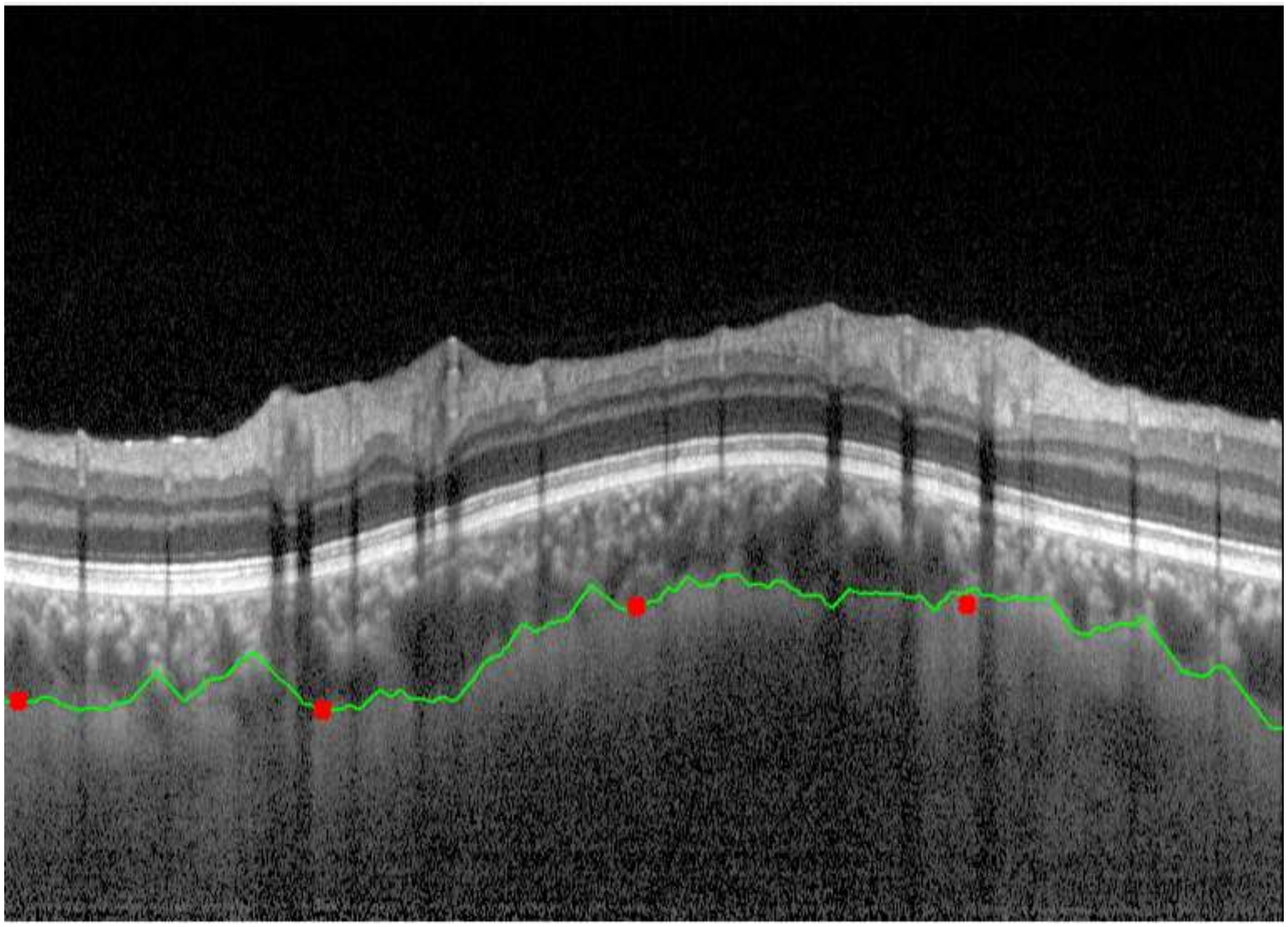

Fig.24. output of the proposed method and maual labeled by the opthalmologist in a circular image

( green line is result of automatic method and red dots are assign with opthalmologist)

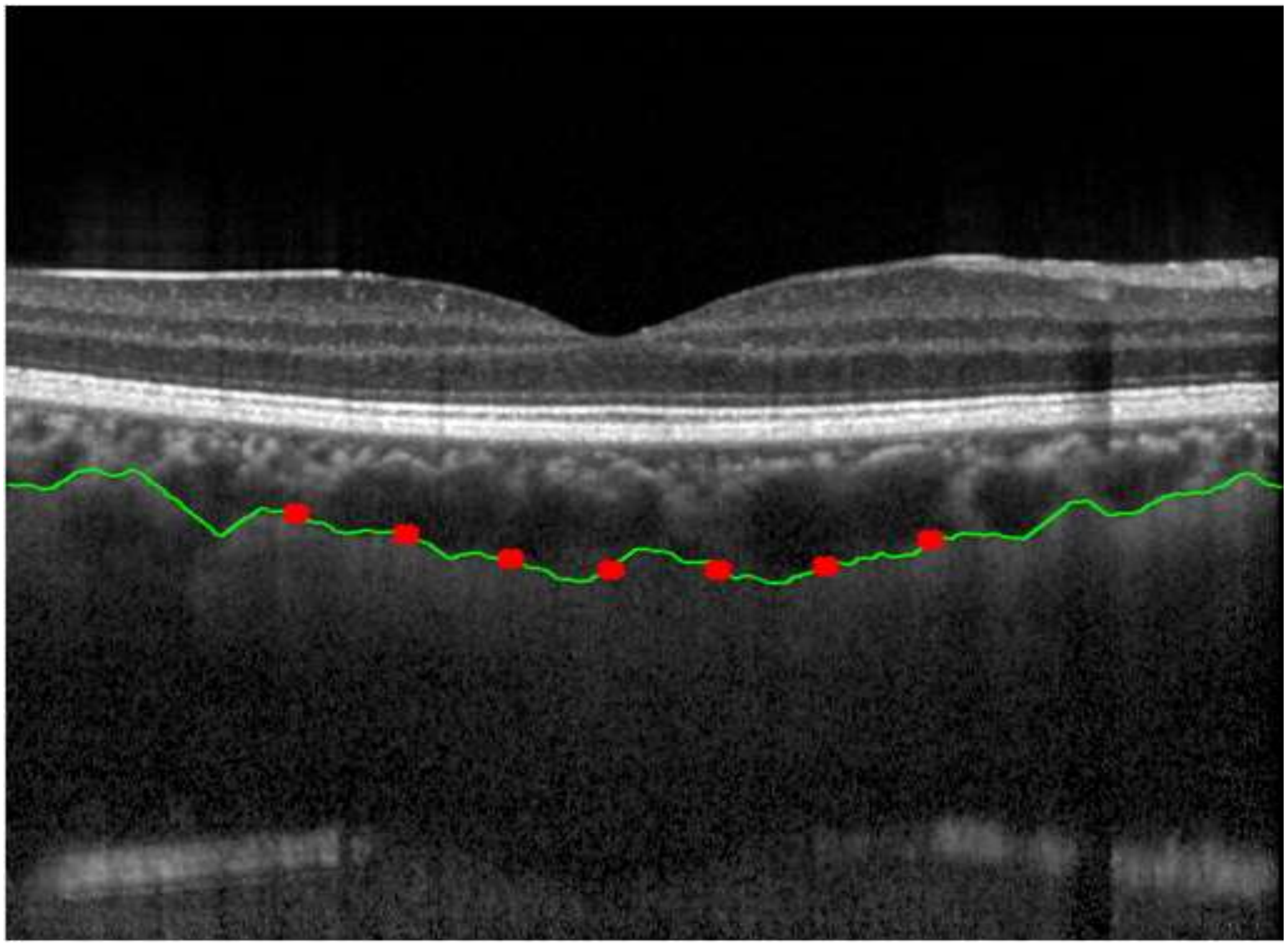

Fig.25. output of the proposed method and maual labeled by the opthalmologist in a macular image

( green line is result of automatic method and red dots are assign with opthalmologist)

| | Error for circular images | Error for macular images |
|---|---|---|
| Error in pixel | 6.55 | 3.34 |
| Error in millimeter | 0.0253 | 0.0129 |

Table.2. Error of the proposed method

## 6.Conclussion

We have proposed an automated, robust and highly accurate method for the detection of the choroid layer. The accuracy can be improved by manual segmentation of the ophthalmologist in desired areas. In our proposed method, which is simple and novel, we apply neutrosophic and graph theory. At first, the image is converted to the neutrosophic space; then, the weight of each two nodes is calculated, and the Dijkstra shortest path algorithm has been employed in order to detect the RPE and choroid layers. Furthermore, the proposed method on prepapillary data is novel and published works are concentrated on macular data and our results has the best accuracy for these images.